\documentclass[sigconf,natbib=true]{acmart}
\usepackage{xspace}
\usepackage{caption}
\usepackage{subcaption}
\usepackage{multirow}
\usepackage[normalem]{ulem}
\usepackage{subcaption}
\usepackage{xcolor}
\usepackage[linesnumbered,ruled]{algorithm2e}
\usepackage{cleveref}
\usepackage{amsmath}
\usepackage{booktabs}
\usepackage{varwidth}
\usepackage{paralist}
\usepackage{wrapfig}
\usepackage{bm}
\usepackage{multirow}
\usepackage[normalem]{ulem}
\useunder{\uline}{\ul}{}
\usepackage{tikz}

\newcommand{\our}{\mbox{\textsc{CaFe}}\xspace}

\usepackage{enumitem}
\AtBeginDocument{%
  \providecommand\BibTeX{{%
    \normalfont B\kern-0.5em{\scshape i\kern-0.25em b}\kern-0.8em\TeX}}}

\setcopyright{acmcopyright}
\copyrightyear{2018}
\acmYear{2018}
\acmDOI{10.1145/1122445.1122456}

\acmConference[Woodstock '18]{Woodstock '18: ACM Symposium on Neural
  Gaze Detection}{June 03--05, 2018}{Woodstock, NY}
\acmBooktitle{Woodstock '18: ACM Symposium on Neural Gaze Detection,
  June 03--05, 2018, Woodstock, NY}
\acmPrice{15.00}
\acmISBN{978-1-4503-XXXX-X/18/06}

\begin{document}



\title{Coarse-to-Fine Sparse Sequential Recommendation}




\author{
  Jiacheng Li$^1$,
  Tong Zhao$^2$, 
  Jin Li$^2$,
  Jim Chan$^2$, 
  Christos Faloutsos$^2$\\
  George Karypis$^2$,
  Soo-Min Pantel$^2$,
  Julian McAuley$^2$
}
\email{
  j9li@eng.ucsd.edu, {zhaoton, jincli, jamchan, faloutso, gkarypis, pantel, jumcaule}@amazon.com
}
\affiliation{%
  \country{$^1$University of California, San Diego \hspace{20pt} $^2$Amazon, United States}
}

\renewcommand{\shortauthors}{Jiacheng and Tong, et al.}

\begin{abstract}
Sequential recommendation aims to model dynamic user behavior from historical interactions. Self-attentive methods have proven effective at capturing short-term dynamics and long-term preferences. Despite their success, these approaches still struggle to model sparse data, 
on which they struggle
to learn high-quality item representations.
We propose to model user dynamics from 
shopping intents and interacted
items simultaneously. The learned 
intents are coarse-grained and work as prior knowledge for item recommendation. To this end, we present a coarse-to-fine self-attention framework, namely \our, which explicitly learns coarse-grained and fine-grained sequential dynamics. Specifically, \our first learns intents from coarse-grained sequences which are dense and hence provide high-quality user intent representations. Then, 
\our fuses intent representations into item encoder outputs to obtain improved item representations. 
Finally, we infer recommended items based on representations of items and corresponding intents.
Experiments on 
sparse datasets show that \our outperforms state-of-the-art self-attentive recommenders by $44.03\%$ NDCG@5 on average.
\end{abstract}




\maketitle

\section{Introduction}
\begin{figure}
    \centering
    \includegraphics[width=\linewidth]{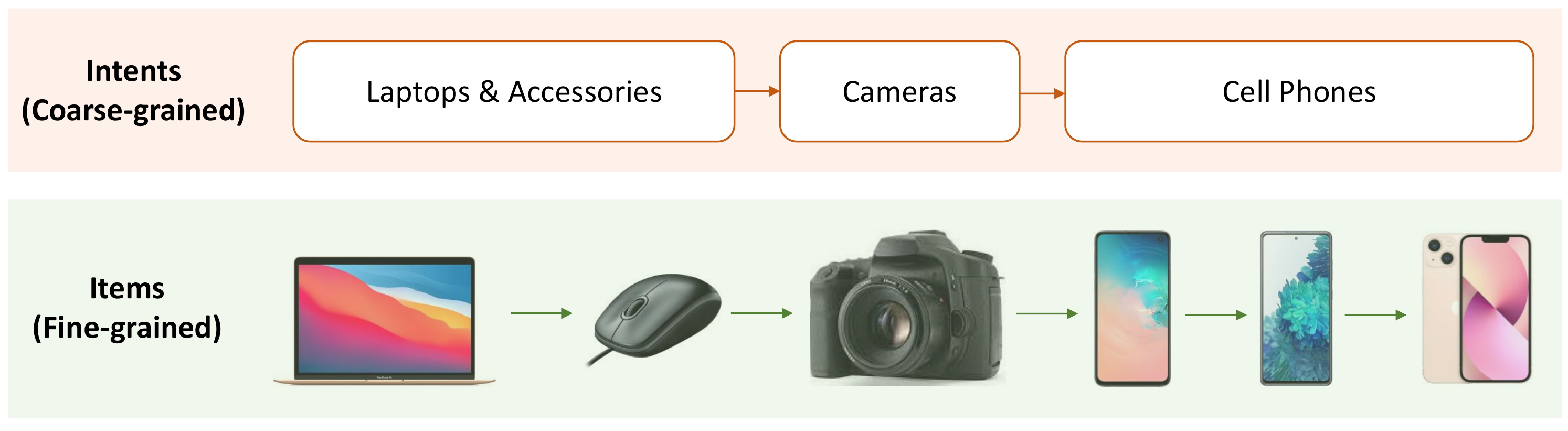}
    \vspace{-2em}
    \caption{Illustration of a coarse-grained sequence (intents) and a fine-grained sequence (items).}
    \label{fig:framework}
\vspace{-2em}
\end{figure}

\begin{figure*}
     \centering
     \begin{subfigure}[b]{0.22\linewidth}
         \centering
         \includegraphics[width=\linewidth]{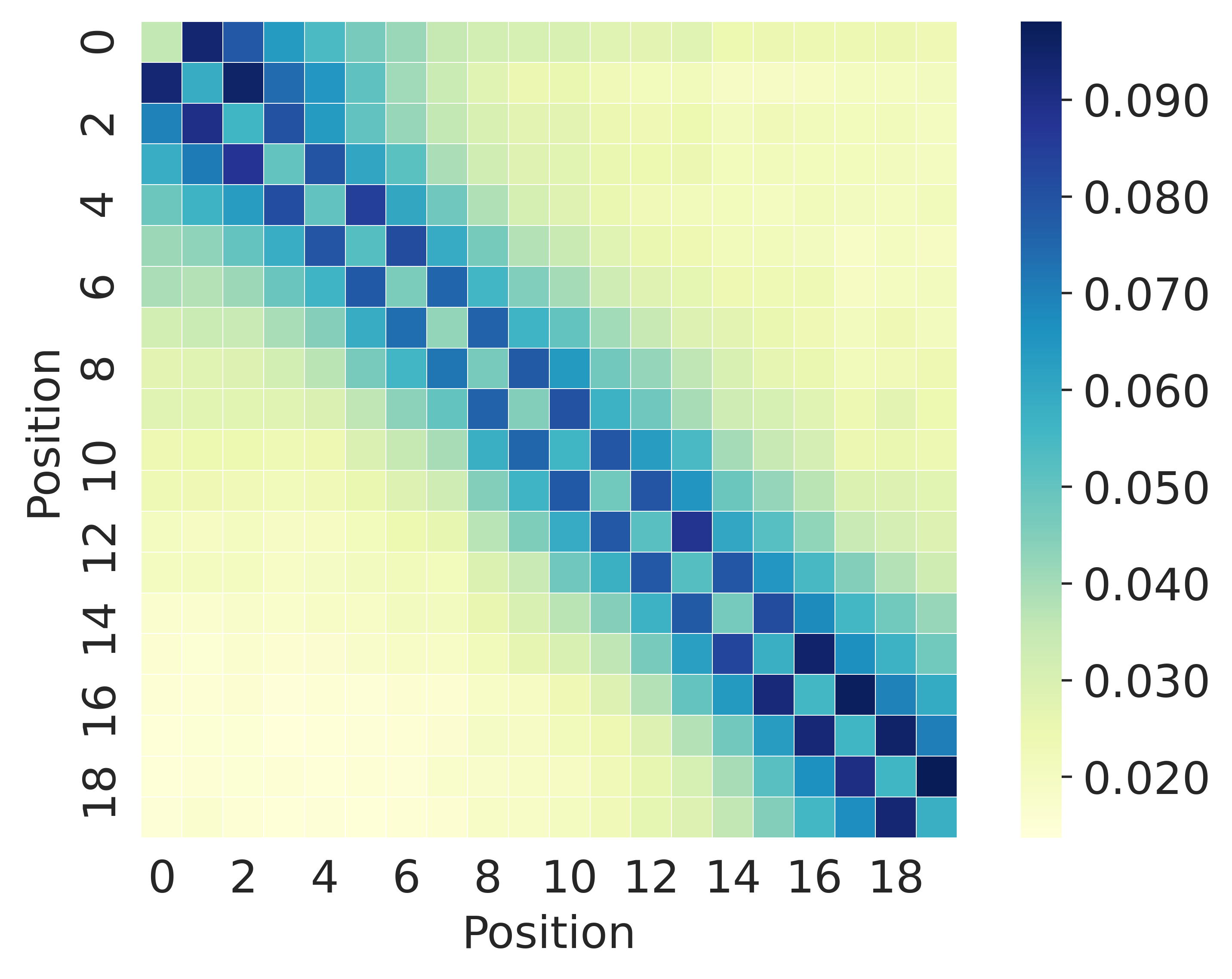}
         \vspace{-2em}
         \caption{\emph{Dense} dataset}
         \label{fig:ml_attn}
     \end{subfigure}
     \hfill
     \begin{subfigure}[b]{0.22\linewidth}
         \centering
         \includegraphics[width=\linewidth]{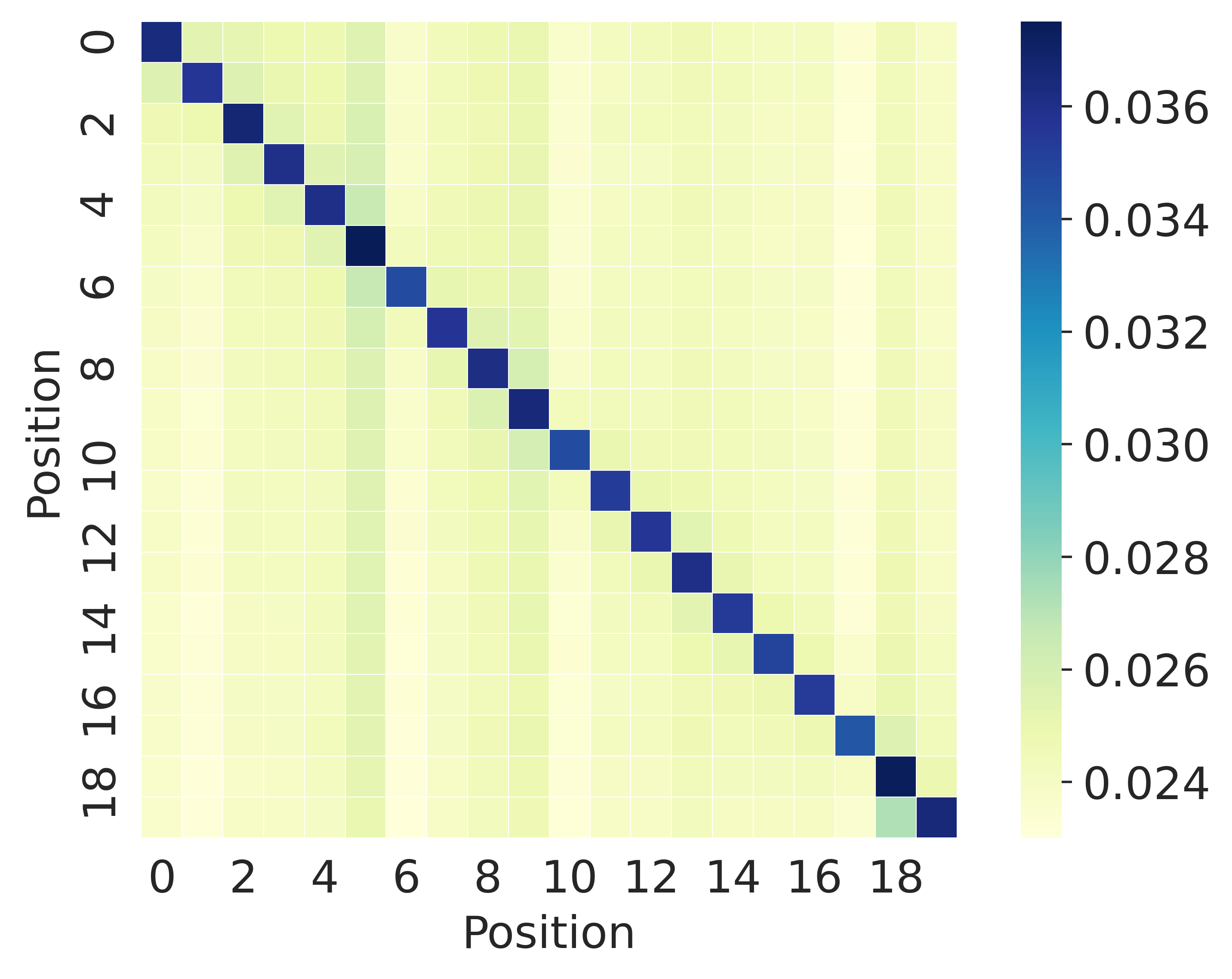}
         \vspace{-2em}
         \caption{\emph{Sparse} dataset}
         \label{fig:amzn_attn}
     \end{subfigure}
     \hfill
     \begin{subfigure}[b]{0.22\linewidth}
         \centering
         \includegraphics[width=\linewidth]{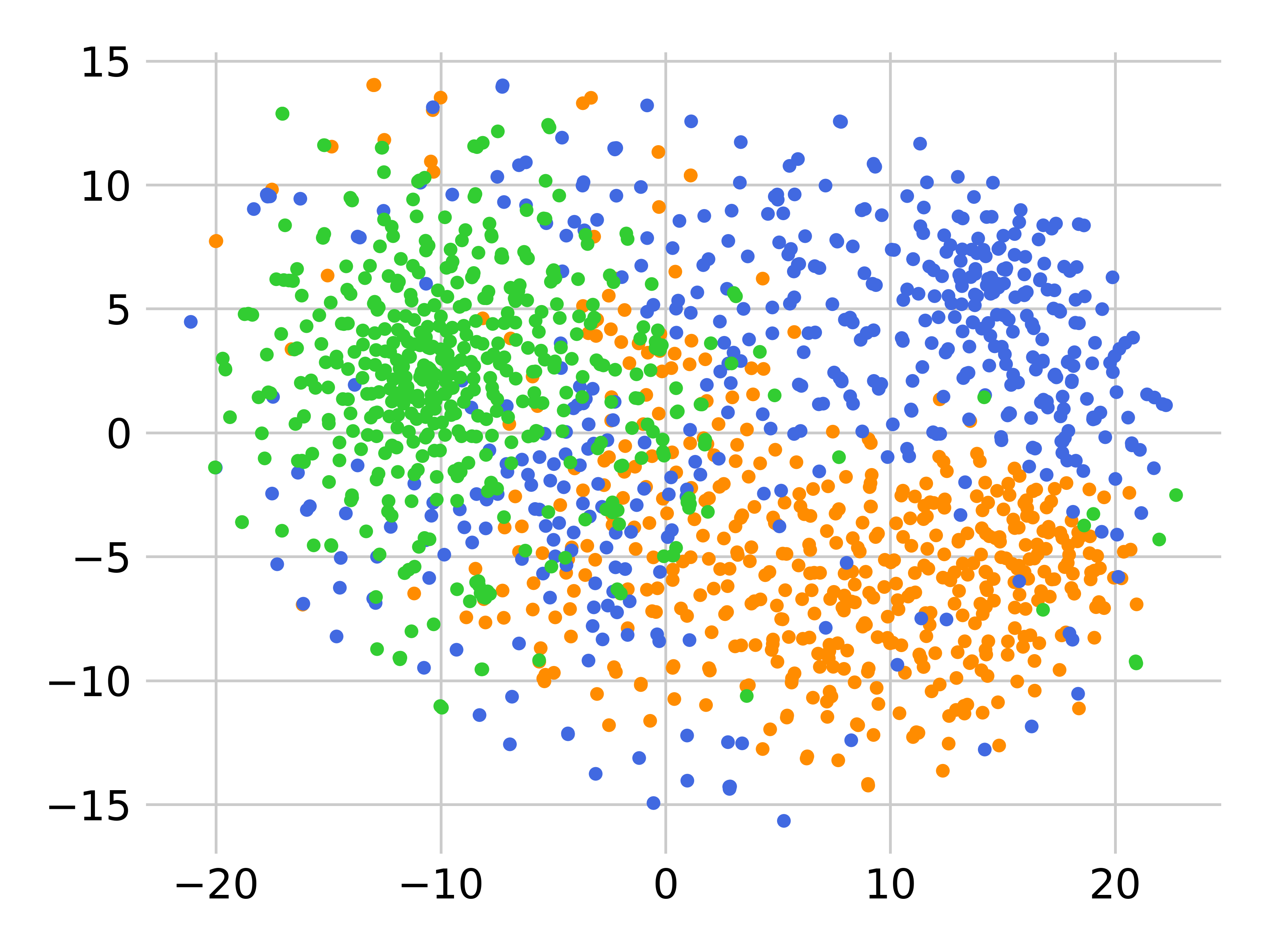}
         \vspace{-2em}
         \caption{Frequent items}
         \label{fig:freq_items}
     \end{subfigure}
     \hfill
     \begin{subfigure}[b]{0.22\linewidth}
         \centering
         \includegraphics[width=\linewidth]{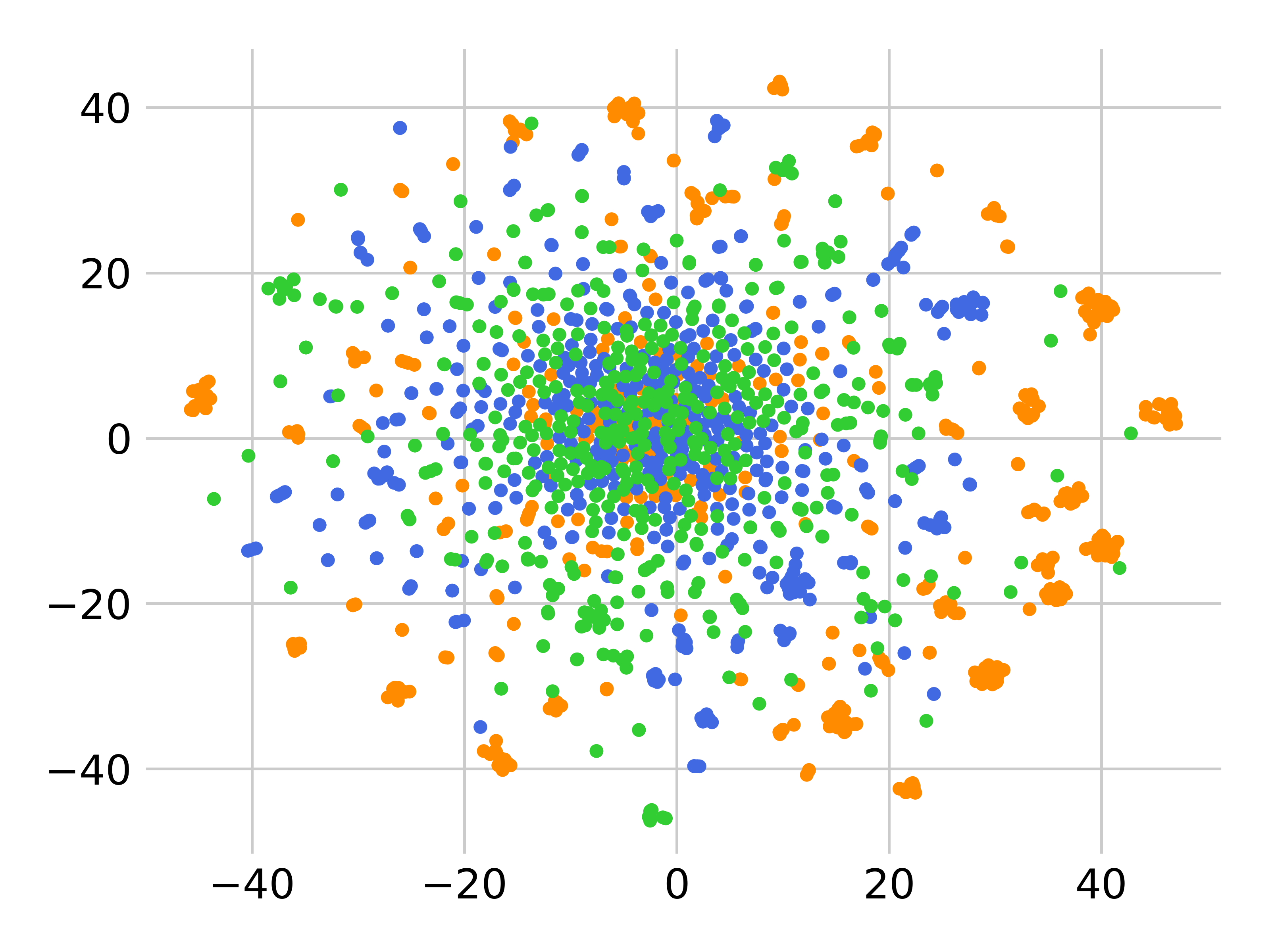}
         \vspace{-2em}
         \caption{Infrequent items}
         \label{fig:infreq_items}
     \end{subfigure}
\vspace{-1em}
        \caption{Motivating experiments. (a) (b) show the average attention map (the first 20 time steps) of BERT4Rec~\cite{Sun2019BERT4RecSR}. (c) (d) show item embeddings (projected by t-SNE~\cite{Maaten2008VisualizingDU}) of BERT4Rec trained on amazon dataset. Item categories are used as labels for coloring (SPORT GOAL-Green; LIP COLOR-Orange; CRIB-Blue).}
        \label{fig:motivate}
\vspace{-1em}
\end{figure*}

The goal of sequential recommender systems is to predict next items for users 
by
modeling historical interactions as temporally-ordered sequences. Sequential recommenders~\cite{Kang2018SelfAttentiveSR, Sun2019BERT4RecSR, Hidasi2016SessionbasedRW, Rendle2010FactorizingPM} capture both \emph{long-term} preferences and \emph{short-term} dynamics of users simultaneously in order to improve recommendation accuracy. 

Previous works employ Markov Chains (MC)~\cite{He2016FusingSM, Rendle2010FactorizingPM}, RNN/CNNs~\cite{Devooght2017LongAS, Li2017NeuralAS, Tang2018PersonalizedTS, Yuan2019ASC} and self-attentive models~\cite{Kang2018SelfAttentiveSR, Sun2019BERT4RecSR, Li2020TimeIA, Wu2020SSEPTSR} for sequential recommendation. Among these approaches, self-attentive recommenders arguably represent the current state-of-the-art, as the self-attention mechanism~\cite{Vaswani2017AttentionIA} is able to efficiently draw context from all past actions 
and obtain short-term dynamics.
Some recent works~\cite{Liu2021NoninvasiveSF, Zhang2019FeaturelevelDS, yan22personalized} incorporate item context features into item representations due to the flexibility of self-attention. Despite the effectiveness of existing self-attentive models, in this paper we argue that sequential recommendation on highly-sparse sequences (i.e.,~containing long-tail items) is still a challenging problem for self-attentive recommenders.

To explore why self-attentive models fail on sparse sequences and validate our motivation, we first conduct two motivating experiments 
(\Cref{sec:moti}) with a representative self-attentive recommender (BERT4Rec~\cite{Sun2019BERT4RecSR}). 
Results reveal that main reasons:
\begin{inparaenum}[(1)]
\item although self-attentive models directly attend on all interactions, they tend to focus on recent items when trained on (item-)sparse datasets.
\item embeddings of long-tail (infrequent) items are under-trained while models represent frequent items well. 
\end{inparaenum}

To address the above problems, we can employ another dense sequence (called an \textit{intent} sequence) to provide prior knowledge and well-trained representations for items. As shown in~\Cref{fig:framework}, although a user interacts with many items (including infrequent items) in the item sequence, several fall under the same shopping intent. For example, the laptop and the mouse belong to the category \texttt{Laptops \& Accessories} 
, and are often purchased together. Hence, if we view categories as intents and explicitly model the intent sequence to predict the next intent, infrequent items can be better understood by their corresponding intent. 
`Intents' in our paper could be categories, taxonomies, or sellers which can reveal high-level `semantics' of items.
Critically,
intent sequences are 
relatively
dense and make it easy to capture long-term preferences of users. Note that some previous works also modeled user shopping intents by implicitly inferring them from items~\cite{Tanjim2020AttentiveSM, Chen2020ImprovingES, Li2021IntentionawareSR} or feature fusion into item representations~\cite{Zhang2019FeaturelevelDS, Liu2021NoninvasiveSF}. However, we find that these \emph{implicit} intent methods do not improve recommendation performance especially on highly-sparse datasets. In contrast, our method \emph{explicitly} learns intent sequences and item sequences which can
improve
sequential recommendation on sparse datasets.

In this work, we propose a \emph{\underline{C}o\underline{a}rse-to-\underline{F}in\underline{e} Framework} (\our), building on self-attentive networks. \our enhances the ability to infrequent item understanding via \emph{explicitly} modeling intent sequences. Specifically, we jointly learn the sequential dynamics of both intents and items with two self-attentive encoders. Compared to previous works that infer the next item via a conditional probability on previous items, \our predicts recommended items based on a joint probability distribution of both items and intents. Experiments show that \our significantly outperforms existing self-attentive recommenders by (on average) $44.03\%$ NDCG@5 on 
sparse datasets.

\section{Preliminaries}
\label{sec:pre}
\subsection{Problem Setup}
We study sequential recommendation for a user set $\mathcal{U}$, an item set $\mathcal{V}$, an intent set $\mathcal{C}$ ($|\mathcal{C}| \ll |\mathcal{V}|$) and a set of user interaction sequences $\mathcal{S} = \{S_1, S_2,\dots,S_{|\mathcal{U}|}\}$. Each item $v\in \mathcal{V}$ has a unique corresponding intent $c\in \mathcal{C}$. A user sequence consists of (temporally-ordered) interactions $S_u = (s^{u}_1, s^{u}_2,\dots,s^{u}_{|S_u|})$, where $S_u\in \mathcal{S}$, $u\in \mathcal{U}$, $s^{u}_i = (v^{u}_i, c^{u}_i)$. Given the interaction history $S_u$, we predict the next item $v^{u}_{|S_u|+1}$.

\subsection{Self-Attentive Recommender}
\label{sec:sasrec}
Self-attentive recommenders~\cite{Kang2018SelfAttentiveSR, Sun2019BERT4RecSR, Wu2020SSEPTSR, Zhang2019FeaturelevelDS} rely on Transformer structure~\cite{Vaswani2017AttentionIA} to encode sequential interactions $\mathcal{S}$. In this paper, our backbone model is a directional self-attentive model SASRec~\cite{Kang2018SelfAttentiveSR}.
\subsubsection{Embedding}
\label{sec:pre_embed}
For an item set $\mathcal{V}$, an embedding table $\mathbf{E}\in \mathbb{R}^{d\times |\mathcal{V}|}$ is used for all items, whose element $\mathbf{e}_i\in \mathbb{R}^d$ denote the embedding for item $v_i$ and $d$ is the latent dimensionality. To be aware of item positions, SASRec maintains a learnable position embedding $\mathbf{P}\in \mathbb{R}^{d\times n}$, where $n$ is the maximum sequence length. All interaction sequences are padded to $n$ with a special `padding' item. Hence, given a padded item sequence $S^v = \{v_1, v_2,\dots,v_n\}$, the input embedding is computed as:
\begin{equation}
\label{eq:embed}
    \mathbf{M}^v = \mathrm{Embedding}(S^v) = [\mathbf{e}_1+\mathbf{p}_1, \mathbf{e}_2+\mathbf{p}_2,\dots, \mathbf{e}_n+\mathbf{p}_n]
\end{equation}
\subsubsection{Transformer Encoder}
The Transformer encoder adopts scaled dot-product attention~\cite{Vaswani2017AttentionIA} denoted as $f_\mathrm{att}$. Given $\mathrm{H}^l_i\in \mathbb{R}^{d}$ is an embedding for $v_i$ after the $l^{\mathrm{th}}$ self-attention layer and $\mathrm{H}^0_i=\mathbf{e}_i+\mathbf{p}_i$, the output from multi-head (\#head=$M$) self-attention is calculated as:
\begin{align}
    \mathbf{O}_i &= \mathrm{Concat}[\mathbf{O}^{(1)}_i,\dots,\mathbf{O}^{(m)}_i,\dots,\mathbf{O}^{(M)}_i]\mathbf{W}_O, \\
    \label{eq:fatt}
    \mathbf{O}^{(m)}_i &= \sum^{n}_{j=1}f_\mathrm{att}(\mathrm{H}^l_i \mathbf{W}^{(m)}_Q, \mathrm{H}^l_j \mathbf{W}^{(m)}_K) \cdot \mathrm{H}^l_j \mathbf{W}^{(m)}_V,
\end{align}
where $\mathbf{W}^{(m)}_Q, \mathbf{W}^{(m)}_K, \mathbf{W}^{(m)}_V \in \mathbb{R}^{d\times d/M}$ are the $m$-th learnable projection matrices; $\mathbf{W}_O \in \mathbb{R}^{d\times d}$ is a learnable matrix to get the output $\mathbf{O}_i$ from concatenated heads. Our backbone SASRec model is a directional self-attention model implemented by forbidding attention weights between $v_i$ and $v_j$ ($j>i$).

To prevent overfitting and achieve a stable training process, the next layer $\mathrm{H}^{l+1}_i$ is generated from $\mathbf{O}_i$ with Residual Connections~\cite{He2016DeepRL}, Layer Normalization~\cite{Ba2016LayerN} and Pointwise Feed-Forward Networks~\cite{Vaswani2017AttentionIA}.

\subsection{Self-Attentive Models on Sparse Data}
\label{sec:moti}
To find reasons that self-attentive models fail on sparse data, we conduct two motivating experiments with BERT4Rec. In experiments, we set hidden size $d=128$ and maximum sequence length of $n=50$. We adopt the same training method as in~\cite{Sun2019BERT4RecSR}.
\subsubsection{Attention Scope}
We investigate the difference between self-attention scope on dense versus sparse data: 
Amazon~\cite{Ni2019JustifyingRU} (av.~2.81 interactions per item) is used as a sparse dataset; A dense version (av.~11.23 interactions per item) is constructed by setting the minimum item frequency to $5$. We visualize the average attention map from the first self-attention layer in~\Cref{fig:ml_attn}/\ref{fig:amzn_attn} which shows that the model attends on more recent items on the sparse dataset, and less recent items for the dense dataset. This indicates that: 
\begin{inparaenum}[(1)]
    \item recent items are important sparse data;
    \item self-attentive models combine long and short-term dynamics, but they still struggle to capture long-term preferences on item-sparse datasets.
\end{inparaenum}
\subsubsection{Trained Embedding}
In this experiment, we explore the difference of trained item embeddings between frequent items and infrequent items in the same dataset. Specifically, we first select three intents (i.e.,~categories in the Amazon dataset), then obtain $500$ the most frequent items and $500$ the most infrequent items in each intent and visualize their trained embeddings in~\Cref{fig:freq_items} and~\ref{fig:infreq_items} respectively. We can see that frequent items with the same intent are usually close to each other (form three colored clusters in~\Cref{fig:freq_items}) while infrequent item embeddings scatter and are mostly around the origin. These observations indicate that 
\begin{inparaenum}[(1)]
    \item the model represents frequent items well, though infrequent items embeddings are under-trained;
    \item intents can provide useful prior knowledge for items because the clusters from well-trained item embeddings are aligned with different intents (see~\Cref{fig:freq_items}). 
\end{inparaenum}

\vspace{-8pt}
\section{Method}
In this section, we propose 
\our to advance sequential recommendation performance on sparse datasets. 

\begin{figure}
    \centering
    \small
    \includegraphics[width=0.8\linewidth]{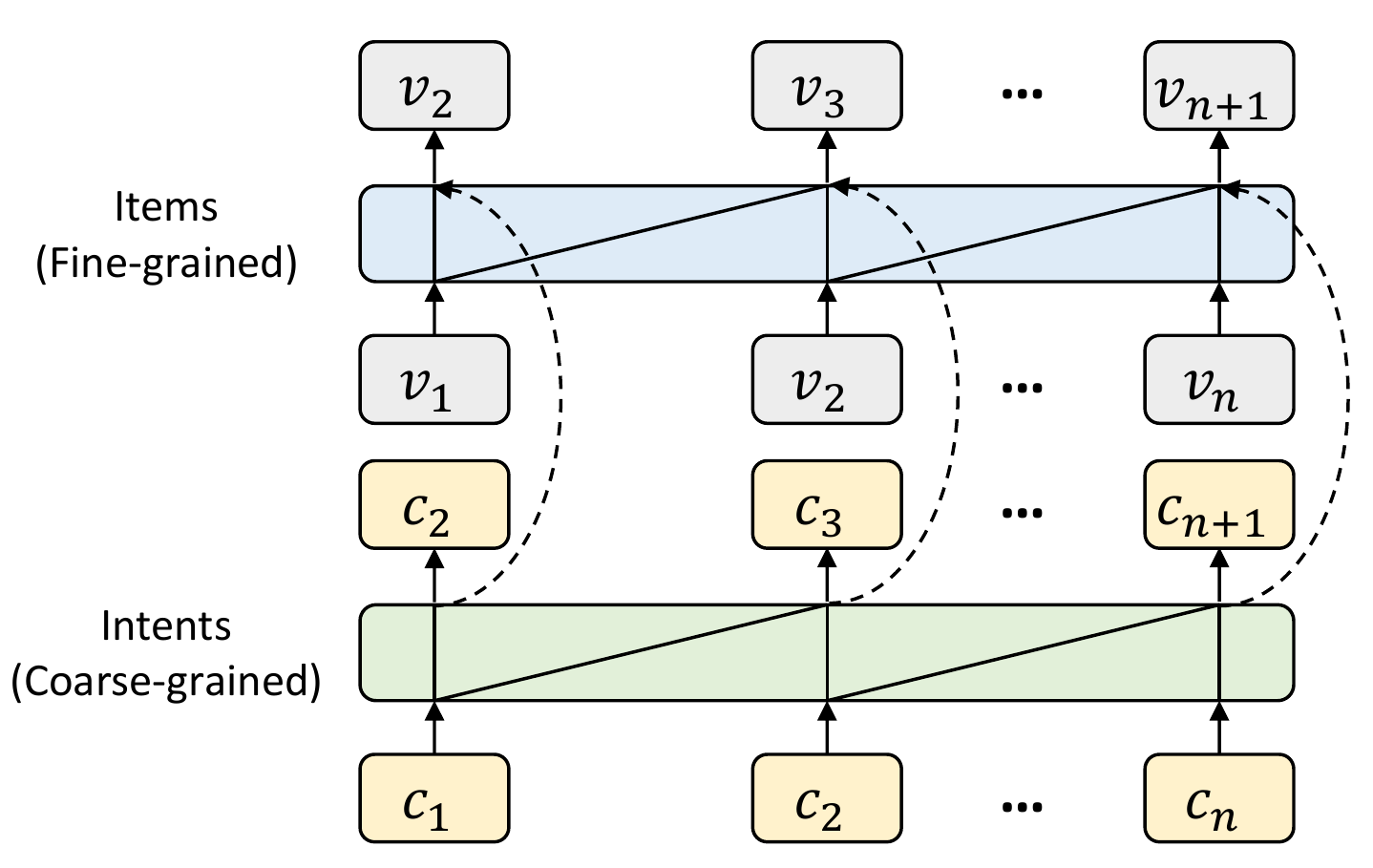}
\vspace{-1em}
    \caption{Framework illustration of \our.}
    \label{fig:method}
\vspace{-2em}
\end{figure}

\vspace{-8pt}
\subsection{Embedding Layer}
In our method, an interaction sequence $S_u$ includes an item sequence $S^v_u$ and an intent sequence $S^c_u$ for user $u$. Hence, We maintain two embedding tables $\mathbf{E}^v\in \mathbb{R}^{d\times |\mathcal{V}|}$ and $\mathbf{E}^c \in \mathbb{R}^{d\times |\mathcal{C}|}$ for items and intents. 
Two separate position embeddings $\mathbf{P^v}\in \mathbb{R}^{d\times n}$ and $\mathbf{P^c}\in \mathbb{R}^{d\times n}$ are created. We encode $S^v_u$ and $S^c_u$ to get input embeddings $\mathbf{M}^v$ and $\mathbf{M}^c$ as follows:
\begin{equation}
    \mathbf{M}^v = \mathrm{Embedding}^v(S^v_u); \mathbf{M}^c = \mathrm{Embedding}^c(S^c_u),
\end{equation}
where $\mathbf{M}^v, \mathbf{M}^c \in \mathbb{R}^{d\times n}$ and $\mathrm{Embedding}(\cdot)$ is the positional embedding operation in~\Cref{eq:embed}.

\vspace{-1em}
\subsection{Coarse-to-Fine Encoder}
Recalling the conclusions in~\Cref{sec:moti}, embeddings of infrequent items are under-trained and the self-attentive model tends to focus on short-term items when the dataset has sparse items. We can also see that intent types are highly aligned with item clusters trained by a self-attentive recommender. Motivated by these observations, we propose to explicitly learn intents in a sequential model and the outputs of the intent model are used as prior knowledge to improve item representations and understand long-term preferences. In this section, we introduce our coarse-to-fine encoder which includes two components, the intent encoder and the item encoder. The overall framework is illustrated in~\Cref{fig:method}.
\subsubsection{Intent Encoder}
For intent sequences, we aim to capture coarse-grained interest dynamics of users. Intent sequences are usually dense because $|\mathcal{C}|$ is much smaller than $|\mathcal{V}|$. Therefore, we apply a standard SASRec model (in~\Cref{sec:sasrec}) as the encoder for intent sequences. Given intent embeddings $\mathbf{M}^c$, outputs of the SASRec encoder are used as intent sequence representations $\mathbf{R}^c \in \mathbb{R}^{d\times n}$.
\subsubsection{Item Encoder}
From our motivating experiments, we see that more recent items are important for next item prediction on sparse datasets. Basically, our item encoder is also a directional Transformer but has enhanced ability to focus on recent items. Inspired by~\cite{He2021LockerLC}, we enhance short-term user dynamics modeling in the item encoder by applying a masking score $\theta_{ij}$ on $f_\mathrm{att}$ in~\Cref{eq:fatt}. Formally, the re-weighted attention weights are calculated by:
\begin{equation}
    f_\mathrm{att}(\mathbf{Q}_i, \mathbf{K}_j) = \frac{\mathrm{exp}(w_{ij})\cdot \theta_{ij}}{\sum^n_{k=1}\mathrm{exp}(w_{ik})\cdot \theta_{ik}}, w_{ij} = \frac{\mathbf{Q}_i \mathbf{K}^T_j}{\sqrt{d}}
\end{equation}
where $\theta = 1$ for a standard scaled dot-product attention and $\sqrt{d}$ is a scale factor. The masking operation $\mathrm{exp}(w_{ij})\cdot \theta_{ij}$ can be rewritten as $\mathrm{exp}(w_{ij}+\mathrm{ln}\theta_{ij})$. We learn $\mathrm{ln}\theta_{ij}$ from $\mathrm{H}^l_i$, $\mathrm{H}^l_j$ and the distance between items $v_i$ and $v_j$:
\begin{equation}
    \mathrm{ln}\theta_{ij} = (\mathrm{H}^l_i \mathbf{W}^{(m)}_Q + \mathrm{H}^l_j \mathbf{W}^{(m)}_K + \mathbf{d}_{ij})\mathbf{W}^{(m)}_L +\mathbf{b}_L
\end{equation}
where $\mathbf{W}^{(m)}_L \in \mathbb{R}^{d/M\times 1}$, $\mathbf{b}_L\in \mathbb{R}^1$, distance embedding $\mathbf{d}_{ij}\in \mathbb{R}^{d/M}$ is the $(n+i-j)$-th vector from distance embedding table $\mathbf{D}\in \mathbb{R}^{d\times 2n}$ and $\mathbf{W}^{(m)}_Q, \mathbf{W}^{(m)}_K$ are from~\Cref{eq:fatt}. We encode $\mathbf{M}^v$ with the item encoder to get the item sequence representations $\mathbf{R}^v\in \mathbb{R}^{d\times n}$

Current item sequence outputs $\mathbf{R}^v$ mostly focus on recent items and cannot represent infrequent items well. To add long preferences and obtain prior knowledge from intents for infrequent items, we add $\mathbf{R}^v$ and $\mathbf{R}^c$ together to get final representations $\mathbf{R}\in \mathbb{R}^{d\times n}$:
\begin{equation}
    \mathbf{R} = \mathbf{R}^v +\mathbf{R}^c
\end{equation}
\subsubsection{Prediction Layer}
In \our, we predict the next intent and item simultaneously from $\mathbf{R}^c$ and $\mathbf{R}$. Specifically, we adopt matrix factorization (MF) to compute the relevance at time step $t$ between encoder outputs and embeddings:
\begin{equation}
    r^c_{j, t} = \mathbf{R}^c_t {\mathbf{E}^c_{j}}^T, r^v_{k, t} = \mathbf{R}_t {\mathbf{E}^v_{k}}^T
\end{equation}
where $\mathbf{E}^c_{j}\in \mathbb{R}^{d}, \mathbf{E}^v_{k}\in \mathbb{R}^{d}$ denotes embeddings of the $j$-th intent and the $k$-th item in $\mathbf{E}^c, \mathbf{E}^v$ respectively.
\vspace{-1em}
\subsection{Network Training}
\our learns from both item sequences and intent sequences, and we adopt the binary cross entropy loss:
\begin{equation}
\small
    \begin{aligned}
    \mathcal{L} =& \mathcal{L}_{c} + \mathcal{L}_v \\
    =&-\sum_{S_u\in \mathcal{S}}\sum_{1\leq t\leq n}\left[\log (\sigma(r^c_{y^c, t}))+\sum_{c_j\not\in S_u}\log (1-\sigma(r^c_{c_j, t}))\right] \\
    &-\sum_{S_u\in \mathcal{S}}\sum_{1\leq t\leq n}\left[\log (\sigma(r^v_{y^v, t}))+\sum_{v_k\not\in S_u}\log (1-\sigma(r^v_{v_k, t}))\right]
    \end{aligned}
\end{equation}
where $y^c, y^v$ are an expected intent and item; $c_j, v_k$ are a negative intent and item randomly generated for each time step in each sequence. Other training details are the same as in SASRec~\cite{Kang2018SelfAttentiveSR}.

\vspace{-1em}
\subsection{Inference}
Previous sequential recommenders infer the next items conditioned on previous items. In contrast, \our computes the joint probability distribution of the item and corresponding intent conditioned on previous intents and items. Formally, at inference we find the item $v_k$ and corresponding intent $c_j$ that maximize the probability:
\begin{equation}
    \begin{aligned}
        P(c_j, v_k|S^c_u, S^v_u, \Theta)
        =& P(c_j|S^c_u, \Theta)P(v_k|c_j, S^c_u, S^v_u, \Theta) \\
        =& \sigma(r^c_{j, t})\sigma(r^v_{k, t})\\
    \end{aligned}
\end{equation}
where $\sigma$ is the sigmoid function, $\Theta$ denotes the parameter set of \our and $S^c_u, S^v_u$ are intent and item sequences for user $u$.

\begin{table}[t]
\centering
\caption{Data statistics.}
\vspace{-10pt}
\scalebox{0.9}{
\small
\setlength{\tabcolsep}{1.2mm}{
\begin{tabular}{@{}lcccccc@{}}
\toprule
Datasets & \#Interaction & \#Item    & \#Intent & \#Sequence & Ave. Length & Density \\ \midrule
Amazon   & 5,370,171     & 1,910,226 & 1,392    & 131,248    & 40.9 & 2e-5  \\
Tmall    & 14,460,516    & 1,788,758 & 9,999    & 131,086    & 110.3 & 6e-5 \\ \bottomrule
\end{tabular}
}}
\vspace{-10pt}
\label{tab:data}
\end{table}

\section{Experiments}
\label{sec:exp}

\begin{table*}[t]
\centering
\small
\caption{Model comparision. The best results are bold and the best baselines are underlined.}
\vspace{-10pt}
\begin{tabular}{@{}llccccccccccc@{}}
\toprule
                        &        & \multicolumn{5}{c}{Item-only Methods}        & \multicolumn{5}{c}{Intent-aware Methods}                                  & \multirow{2}{*}{Improvement} \\ \cmidrule(lr){3-7} \cmidrule(lr){8-12}
Dataset                 & Metric & PopRec & SASRec & BERT4Rec & SSE-PT & LOCKER & NOVA   & FDSA   & BERT-F & \multicolumn{1}{l}{LOCKER-F} & \our            &                              \\ \midrule
\multirow{3}{*}{Amazon} & NDCG@5 & 0.0286 & 0.1418 & 0.1830   & 0.2108 & 0.2170 & 0.0281 & 0.0670 & 0.2199 & {\ul 0.2436}                 & \textbf{0.3733} & +53.24\%                     \\
                        & HR@5   & 0.0487 & 0.1844 & 0.2240   & 0.2501 & 0.2597 & 0.0475 & 0.1089 & 0.2676 & {\ul 0.2947}                 & \textbf{0.4813} & +63.32\%                     \\
                        & MRR    & 0.0485 & 0.1522 & 0.1956   & 0.2239 & 0.2297 & 0.0477 & 0.0857 & 0.2329 & {\ul 0.2529}                 & \textbf{0.3656} & +44.56\%                     \\ \midrule
\multirow{3}{*}{Tmall}  & NDCG@5 & 0.0360 & 0.0741 & 0.2753   & 0.2106 & 0.2961 & 0.0501 & 0.1083 & 0.2998 & {\ul 0.3182}                 & \textbf{0.4290} & +34.82\%                     \\
                        & HR@5   & 0.0596 & 0.1205 & 0.3673   & 0.2977 & 0.3872 & 0.0812 & 0.1685 & 0.3917 & {\ul 0.4098}                 & \textbf{0.5152} & +25.72\%                     \\
                        & MRR    & 0.0577 & 0.0948 & 0.2782   & 0.2173 & 0.2979 & 0.0716 & 0.1265 & 0.3014 & {\ul 0.3189}                 & \textbf{0.4268} & +33.84\%                     \\ \bottomrule
\end{tabular}
\label{tab:performance}
\vspace{-1em}
\end{table*}

\subsection{Experimental Setting}
\subsubsection{Data}
We consider two sparse datasets (see~\Cref{tab:data}): \textbf{Amazon}~\cite{Ni2019JustifyingRU} is collected from Amazon.com, and we use item categories as 
coarse-grained sequences; \textbf{Tmall} is released in the IJCAI-15 challenge~\cite{tmalldata}. Sellers of products are used as 
coarse-grained sequences. We follow~\cite{Kang2018SelfAttentiveSR, Sun2019BERT4RecSR} to conduct a leave-last-2-out data split.

\subsubsection{Baselines}
We compare two groups of works as our baselines which include methods with only items and methods using both intents and items.
\emph{Item-only Methods}:
\textbf{PopRec}, a baseline method that recommends items according to item occurrences in the dataset. 
\textbf{SASRec}~\cite{Kang2018SelfAttentiveSR}, a directional self-attention method that is used as our backbone model.
\textbf{BERT4Rec}~\cite{Sun2019BERT4RecSR}, a bi-directional self-attention method that learns to recommend items via a cloze task similar to BERT~\cite{Devlin2019BERTPO}.
\textbf{SSE-PT}~\cite{Wu2020SSEPTSR}, extends SASRec by introducing explicit user representations.
\textbf{LOCKER}~\cite{He2021LockerLC}, enhances short-term user dynamics via local self-attention.
\emph{Intent-aware Methods}:
\textbf{NOVA}~\cite{Liu2021NoninvasiveSF}, uses non-invasive self-attention to leverage side information. We use intents as side information.
\textbf{FDSA}~\cite{Zhang2019FeaturelevelDS}, applies separated item and feature sequences but does not explicitly learn the feature sequences.
\textbf{BERT-F, LOCKER-F}, our extension of BERT4Rec and LOCKER, which incorporate intents in the same way as FDSA.

\subsubsection{Evaluation and Implementation}
We choose truncated Hit Ratio (HR@K), Normalized Discounted Cumulative Gain (NDCG) and Mean Reciprocal Rank (MRR) to measure ranking quality (K=5). Following BERT4Rec~\cite{Sun2019BERT4RecSR}, we randomly sample $100$ negative items according to their popularity for each ground truth item. For all baselines on two datasets, the maximum length of sequences $n$ is $50$; hidden size $d$ is $128$; batch size is $64$. We implement all models and tune other hyper-parameters following authors' suggestions.
\vspace{-1em}
\subsection{Result Analysis}
\subsubsection{General Performance}
\Cref{tab:performance} shows ranking performance on two datasets. We find that:
\begin{inparaenum}[(1)]
\item Previous intent-aware methods that fuse intent features into item representations achieve limited improvements. We think the reason is that baselines did not learn intent representations from intent sequences but from item sequences. However, on such sparse datasets, items hardly provide supervision for intent learning hence low-quality intent features cannot provide useful information or even result in performance decay (NOVA and FDSA).
\item Compared to global attention (BERT4Rec), local attention (LOCKER) can consistently improve recommendations on these two datasets which validate our observations in motivating experiments.
\item \our outperforms all baselines significantly. Compared to the strongest baseline LOCKER-F, our model gains about $44.03\%$ NDCG@5 and about $44.53\%$ HR@5 improvements on average. 
See~\Cref{sec:ablation} for detailed analysis.
\end{inparaenum}

\begin{table}[t]
\centering
\scalebox{0.97}{
\small
\setlength{\tabcolsep}{1.0mm}{
\begin{tabular}{@{}lclcccc@{}}
\toprule
       & \begin{tabular}[c]{@{}c@{}}Backbone\\ (SASRec)\end{tabular} & \multicolumn{1}{c}{\begin{tabular}[c]{@{}c@{}}+(1)\\ (FDSA)\end{tabular}} & +(1)(2)   & +(1)(2)(3) & +(1)(2)(4) & \begin{tabular}[c]{@{}c@{}}+(1)(2)(3)(4)\\ (\our)\end{tabular} \\ \midrule
NDCG@5 & 0.0741                                                      & 0.1083                                                                    & 0.3045 & 0.3159  & 0.4254  & \textbf{0.4290}                                             \\
HR@5   & 0.1205                                                      & 0.1685                                                                    & 0.3938 & 0.4066  & 0.5117  & \textbf{0.5152}                                             \\
MRR    & 0.0948                                                      & 0.1265                                                                    & 0.3069 & 0.3172  & 0.4239  & \textbf{0.4268}                                             \\ \bottomrule
\end{tabular}
}}
\caption{Ablation study on Tmall dataset. (1) fusing intents into item embeddings; (2) modeling intents explicitly; (3) local self-attention of item encoder; (4) inference with joint probability distribution of items and corresponding intents.}
\label{tab:ablation}
\vspace{-3em}
\end{table}

\subsubsection{Ablation Study}
\label{sec:ablation}
To validate the effectiveness of our proposed method, we conduct an ablation study on Tmall dataset;~\Cref{tab:ablation} shows results. Compared to (1) FDSA which fuses intent features into item representations, (2) modeling intents explicitly (i.e.,~learning intent representations from intent sequences) is critical to make intent representations effective for items. (4) joint probability inference largely improves recommendation performance by providing coarse-grained knowledge during inference. (3) local self-attention can further improve results by focusing on more recent items. 

\subsubsection{Improvement vs.~Sparsity}
\begin{figure}[t]
    \centering
    \includegraphics[width=0.75\linewidth]{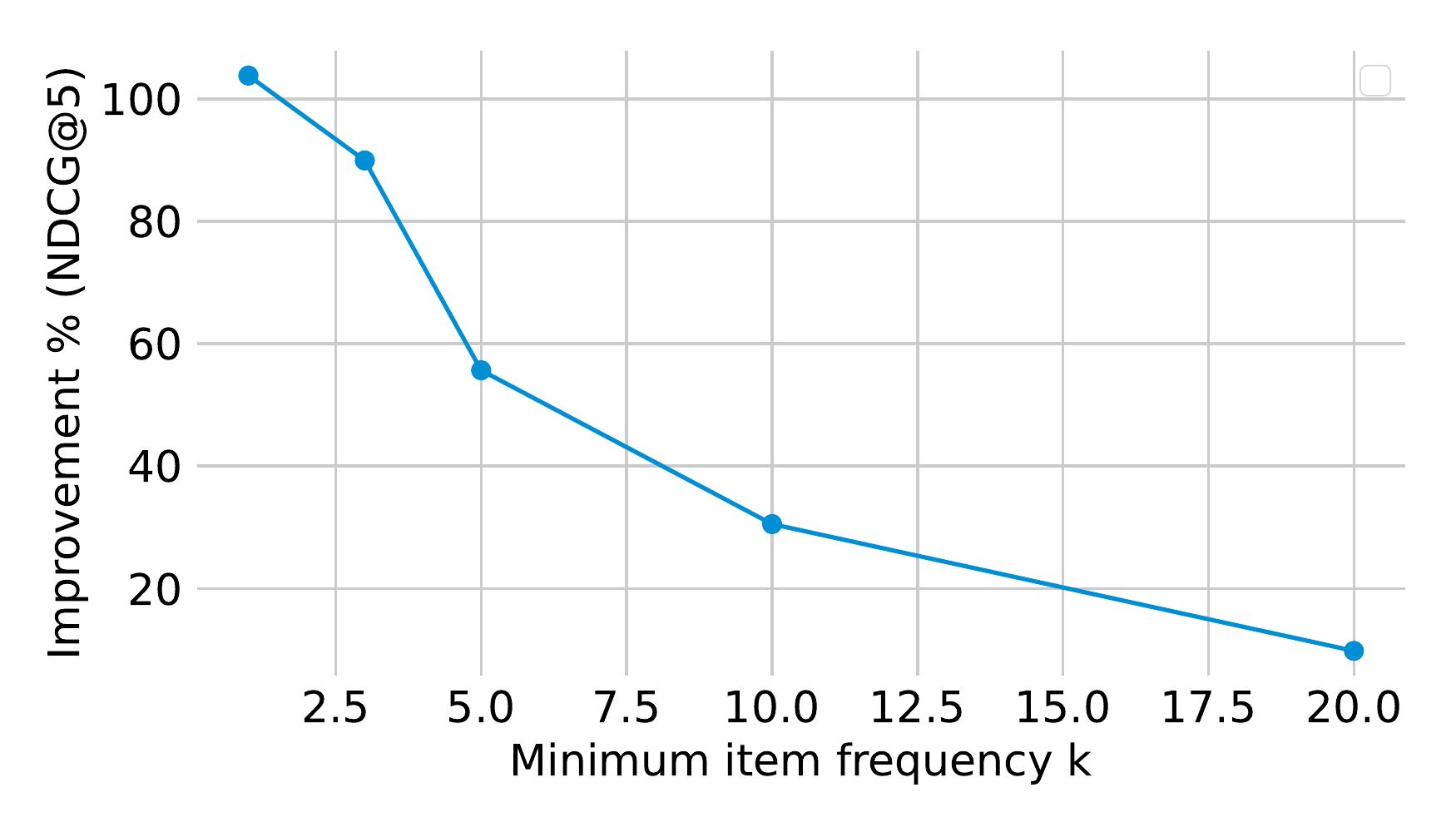}
\vspace{-2em}
    \caption{Improvement on Amazon compared to BERT4Rec.}
    \label{fig:sparsity}
\vspace{-1em}
\end{figure}
We investigate \our performance compared to BERT4Rec with different dataset sparsity in~\Cref{fig:sparsity}. Models (\our and BERT4Rec) are trained and tested on Amazon datasets with different minimum item frequency $k$.
Smaller $k$ means the dataset is sparser. We can see that the improvement is more than $100\%$ on the dataset with $k=1$ (original) and the improvement is less than $10\%$ on the dataset with $k=20$ (the most dense). The results show the effectiveness of \our on sparse datasets.


\section{Conclusion}
Self-attentive recommenders have shown promising results in sequential recommendation. However, we find that existing methods still struggle to learn high-quality item representations from sparse data. In this paper, we introduce a coarse-to-fine framework (\our) that explicitly models intent sequences and enhances infrequent item representations by knowledge from intents. Furthermore, we propose to infer recommended items based on joint probability of intents and items. Experimental results show that \our significantly improves recommendation performance on sparse datasets.


\clearpage

\bibliographystyle{ACM-Reference-Format}
\bibliography{ref}


\begin{thebibliography}{25}


\ifx \showCODEN    \undefined \def \showCODEN     #1{\unskip}     \fi
\ifx \showDOI      \undefined \def \showDOI       #1{#1}\fi
\ifx \showISBNx    \undefined \def \showISBNx     #1{\unskip}     \fi
\ifx \showISBNxiii \undefined \def \showISBNxiii  #1{\unskip}     \fi
\ifx \showISSN     \undefined \def \showISSN      #1{\unskip}     \fi
\ifx \showLCCN     \undefined \def \showLCCN      #1{\unskip}     \fi
\ifx \shownote     \undefined \def \shownote      #1{#1}          \fi
\ifx \showarticletitle \undefined \def \showarticletitle #1{#1}   \fi
\ifx \showURL      \undefined \def \showURL       {\relax}        \fi
\providecommand\bibfield[2]{#2}
\providecommand\bibinfo[2]{#2}
\providecommand\natexlab[1]{#1}
\providecommand\showeprint[2][]{arXiv:#2}

\bibitem[\protect\citeauthoryear{??}{tma}{2015}]%
        {tmalldata}
 \bibinfo{year}{2015}\natexlab{}.
\newblock
  \showarticletitle{https://ijcai-15.org/repeat-buyers-prediction-competition/}.
  In \bibinfo{booktitle}{\emph{IJCAI}}.
\newblock


\bibitem[\protect\citeauthoryear{Ba, Kiros, and Hinton}{Ba
  et~al\mbox{.}}{2016}]%
        {Ba2016LayerN}
\bibfield{author}{\bibinfo{person}{Jimmy Ba}, \bibinfo{person}{Jamie~Ryan
  Kiros}, {and} \bibinfo{person}{Geoffrey~E. Hinton}.}
  \bibinfo{year}{2016}\natexlab{}.
\newblock \showarticletitle{Layer Normalization}.
\newblock \bibinfo{journal}{\emph{ArXiv}}  \bibinfo{volume}{abs/1607.06450}
  (\bibinfo{year}{2016}).
\newblock


\bibitem[\protect\citeauthoryear{Chen, Ren, Cai, Sun, and de~Rijke}{Chen
  et~al\mbox{.}}{2020}]%
        {Chen2020ImprovingES}
\bibfield{author}{\bibinfo{person}{Wanyu Chen}, \bibinfo{person}{Pengjie Ren},
  \bibinfo{person}{Fei Cai}, \bibinfo{person}{Fei Sun}, {and}
  \bibinfo{person}{M. de Rijke}.} \bibinfo{year}{2020}\natexlab{}.
\newblock \showarticletitle{Improving End-to-End Sequential Recommendations
  with Intent-aware Diversification}.
\newblock \bibinfo{journal}{\emph{Proceedings of the 29th ACM International
  Conference on Information \& Knowledge Management}} (\bibinfo{year}{2020}).
\newblock


\bibitem[\protect\citeauthoryear{Devlin, Chang, Lee, and Toutanova}{Devlin
  et~al\mbox{.}}{2019}]%
        {Devlin2019BERTPO}
\bibfield{author}{\bibinfo{person}{Jacob Devlin}, \bibinfo{person}{Ming-Wei
  Chang}, \bibinfo{person}{Kenton Lee}, {and} \bibinfo{person}{Kristina
  Toutanova}.} \bibinfo{year}{2019}\natexlab{}.
\newblock \showarticletitle{BERT: Pre-training of Deep Bidirectional
  Transformers for Language Understanding}.
\newblock \bibinfo{journal}{\emph{ArXiv}}  \bibinfo{volume}{abs/1810.04805}
  (\bibinfo{year}{2019}).
\newblock


\bibitem[\protect\citeauthoryear{Devooght and Bersini}{Devooght and
  Bersini}{2017}]%
        {Devooght2017LongAS}
\bibfield{author}{\bibinfo{person}{Robin Devooght} {and}
  \bibinfo{person}{Hugues Bersini}.} \bibinfo{year}{2017}\natexlab{}.
\newblock \showarticletitle{Long and Short-Term Recommendations with Recurrent
  Neural Networks}.
\newblock \bibinfo{journal}{\emph{Proceedings of the 25th Conference on User
  Modeling, Adaptation and Personalization}} (\bibinfo{year}{2017}).
\newblock


\bibitem[\protect\citeauthoryear{He, Zhang, Ren, and Sun}{He
  et~al\mbox{.}}{2016}]%
        {He2016DeepRL}
\bibfield{author}{\bibinfo{person}{Kaiming He}, \bibinfo{person}{X. Zhang},
  \bibinfo{person}{Shaoqing Ren}, {and} \bibinfo{person}{Jian Sun}.}
  \bibinfo{year}{2016}\natexlab{}.
\newblock \showarticletitle{Deep Residual Learning for Image Recognition}.
\newblock \bibinfo{journal}{\emph{2016 IEEE Conference on Computer Vision and
  Pattern Recognition (CVPR)}} (\bibinfo{year}{2016}),
  \bibinfo{pages}{770--778}.
\newblock


\bibitem[\protect\citeauthoryear{He and McAuley}{He and McAuley}{2016}]%
        {He2016FusingSM}
\bibfield{author}{\bibinfo{person}{Ruining He} {and} \bibinfo{person}{Julian
  McAuley}.} \bibinfo{year}{2016}\natexlab{}.
\newblock \showarticletitle{Fusing Similarity Models with Markov Chains for
  Sparse Sequential Recommendation}.
\newblock \bibinfo{journal}{\emph{2016 IEEE 16th International Conference on
  Data Mining (ICDM)}} (\bibinfo{year}{2016}), \bibinfo{pages}{191--200}.
\newblock


\bibitem[\protect\citeauthoryear{He, Zhao, Lin, Wang, Kale, and McAuley}{He
  et~al\mbox{.}}{2021}]%
        {He2021LockerLC}
\bibfield{author}{\bibinfo{person}{Zhankui He}, \bibinfo{person}{Handong Zhao},
  \bibinfo{person}{Zhe Lin}, \bibinfo{person}{Zhaowen Wang},
  \bibinfo{person}{Ajinkya Kale}, {and} \bibinfo{person}{Julian McAuley}.}
  \bibinfo{year}{2021}\natexlab{}.
\newblock \showarticletitle{Locker: Locally Constrained Self-Attentive
  Sequential Recommendation}.
\newblock \bibinfo{journal}{\emph{Proceedings of the 30th ACM International
  Conference on Information \& Knowledge Management}} (\bibinfo{year}{2021}).
\newblock


\bibitem[\protect\citeauthoryear{Hidasi, Karatzoglou, Baltrunas, and
  Tikk}{Hidasi et~al\mbox{.}}{2016}]%
        {Hidasi2016SessionbasedRW}
\bibfield{author}{\bibinfo{person}{Bal{\'a}zs Hidasi},
  \bibinfo{person}{Alexandros Karatzoglou}, \bibinfo{person}{Linas Baltrunas},
  {and} \bibinfo{person}{Domonkos Tikk}.} \bibinfo{year}{2016}\natexlab{}.
\newblock \showarticletitle{Session-based Recommendations with Recurrent Neural
  Networks}.
\newblock \bibinfo{journal}{\emph{CoRR}}  \bibinfo{volume}{abs/1511.06939}
  (\bibinfo{year}{2016}).
\newblock


\bibitem[\protect\citeauthoryear{Kang and McAuley}{Kang and McAuley}{2018}]%
        {Kang2018SelfAttentiveSR}
\bibfield{author}{\bibinfo{person}{Wang-Cheng Kang} {and}
  \bibinfo{person}{Julian McAuley}.} \bibinfo{year}{2018}\natexlab{}.
\newblock \showarticletitle{Self-Attentive Sequential Recommendation}.
\newblock \bibinfo{journal}{\emph{2018 IEEE International Conference on Data
  Mining (ICDM)}} (\bibinfo{year}{2018}), \bibinfo{pages}{197--206}.
\newblock


\bibitem[\protect\citeauthoryear{Li, Wang, Zhang, Ma, Cui, and Zhu}{Li
  et~al\mbox{.}}{2021}]%
        {Li2021IntentionawareSR}
\bibfield{author}{\bibinfo{person}{Haoyang Li}, \bibinfo{person}{Xin Wang},
  \bibinfo{person}{Ziwei Zhang}, \bibinfo{person}{Jianxin Ma},
  \bibinfo{person}{Peng Cui}, {and} \bibinfo{person}{Wenwu Zhu}.}
  \bibinfo{year}{2021}\natexlab{}.
\newblock \showarticletitle{Intention-aware Sequential Recommendation with
  Structured Intent Transition}.
\newblock \bibinfo{journal}{\emph{IEEE Transactions on Knowledge and Data
  Engineering}} (\bibinfo{year}{2021}), \bibinfo{pages}{1--1}.
\newblock


\bibitem[\protect\citeauthoryear{Li, Ren, Chen, Ren, Lian, and Ma}{Li
  et~al\mbox{.}}{2017}]%
        {Li2017NeuralAS}
\bibfield{author}{\bibinfo{person}{Jing Li}, \bibinfo{person}{Pengjie Ren},
  \bibinfo{person}{Zhumin Chen}, \bibinfo{person}{Zhaochun Ren},
  \bibinfo{person}{Tao Lian}, {and} \bibinfo{person}{Jun Ma}.}
  \bibinfo{year}{2017}\natexlab{}.
\newblock \showarticletitle{Neural Attentive Session-based Recommendation}.
\newblock \bibinfo{journal}{\emph{Proceedings of the 2017 ACM on Conference on
  Information and Knowledge Management}} (\bibinfo{year}{2017}).
\newblock


\bibitem[\protect\citeauthoryear{Li, Wang, and McAuley}{Li
  et~al\mbox{.}}{2020}]%
        {Li2020TimeIA}
\bibfield{author}{\bibinfo{person}{Jiacheng Li}, \bibinfo{person}{Yujie Wang},
  {and} \bibinfo{person}{Julian McAuley}.} \bibinfo{year}{2020}\natexlab{}.
\newblock \showarticletitle{Time Interval Aware Self-Attention for Sequential
  Recommendation}.
\newblock \bibinfo{journal}{\emph{Proceedings of the 13th International
  Conference on Web Search and Data Mining}} (\bibinfo{year}{2020}).
\newblock


\bibitem[\protect\citeauthoryear{Liu, Li, Cai, Dong, Zhu, and Shang}{Liu
  et~al\mbox{.}}{2021}]%
        {Liu2021NoninvasiveSF}
\bibfield{author}{\bibinfo{person}{Chang Liu}, \bibinfo{person}{Xiaoguang Li},
  \bibinfo{person}{Guohao Cai}, \bibinfo{person}{Zhenhua Dong},
  \bibinfo{person}{Hong Zhu}, {and} \bibinfo{person}{Lifeng Shang}.}
  \bibinfo{year}{2021}\natexlab{}.
\newblock \showarticletitle{Non-invasive Self-attention for Side Information
  Fusion in Sequential Recommendation}. In \bibinfo{booktitle}{\emph{AAAI}}.
\newblock


\bibitem[\protect\citeauthoryear{Ni, Li, and McAuley}{Ni et~al\mbox{.}}{2019}]%
        {Ni2019JustifyingRU}
\bibfield{author}{\bibinfo{person}{Jianmo Ni}, \bibinfo{person}{Jiacheng Li},
  {and} \bibinfo{person}{Julian McAuley}.} \bibinfo{year}{2019}\natexlab{}.
\newblock \showarticletitle{Justifying Recommendations using Distantly-Labeled
  Reviews and Fine-Grained Aspects}. In \bibinfo{booktitle}{\emph{EMNLP}}.
\newblock


\bibitem[\protect\citeauthoryear{Rendle, Freudenthaler, and
  Schmidt-Thieme}{Rendle et~al\mbox{.}}{2010}]%
        {Rendle2010FactorizingPM}
\bibfield{author}{\bibinfo{person}{Steffen Rendle}, \bibinfo{person}{Christoph
  Freudenthaler}, {and} \bibinfo{person}{Lars Schmidt-Thieme}.}
  \bibinfo{year}{2010}\natexlab{}.
\newblock \showarticletitle{Factorizing personalized Markov chains for
  next-basket recommendation}. In \bibinfo{booktitle}{\emph{WWW '10}}.
\newblock


\bibitem[\protect\citeauthoryear{Sun, Liu, Wu, Pei, Lin, Ou, and Jiang}{Sun
  et~al\mbox{.}}{2019}]%
        {Sun2019BERT4RecSR}
\bibfield{author}{\bibinfo{person}{Fei Sun}, \bibinfo{person}{Jun Liu},
  \bibinfo{person}{Jian Wu}, \bibinfo{person}{Changhua Pei},
  \bibinfo{person}{Xiao Lin}, \bibinfo{person}{Wenwu Ou}, {and}
  \bibinfo{person}{Peng Jiang}.} \bibinfo{year}{2019}\natexlab{}.
\newblock \showarticletitle{BERT4Rec: Sequential Recommendation with
  Bidirectional Encoder Representations from Transformer}.
\newblock \bibinfo{journal}{\emph{Proceedings of the 28th ACM International
  Conference on Information and Knowledge Management}} (\bibinfo{year}{2019}).
\newblock


\bibitem[\protect\citeauthoryear{Tang and Wang}{Tang and Wang}{2018}]%
        {Tang2018PersonalizedTS}
\bibfield{author}{\bibinfo{person}{Jiaxi Tang} {and} \bibinfo{person}{Ke
  Wang}.} \bibinfo{year}{2018}\natexlab{}.
\newblock \showarticletitle{Personalized Top-N Sequential Recommendation via
  Convolutional Sequence Embedding}.
\newblock \bibinfo{journal}{\emph{Proceedings of the Eleventh ACM International
  Conference on Web Search and Data Mining}} (\bibinfo{year}{2018}).
\newblock


\bibitem[\protect\citeauthoryear{Tanjim, Su, Benjamin, Hu, Hong, and
  McAuley}{Tanjim et~al\mbox{.}}{2020}]%
        {Tanjim2020AttentiveSM}
\bibfield{author}{\bibinfo{person}{Md.~Mehrab Tanjim}, \bibinfo{person}{Congzhe
  Su}, \bibinfo{person}{Ethan Benjamin}, \bibinfo{person}{Diane~J. Hu},
  \bibinfo{person}{Liangjie Hong}, {and} \bibinfo{person}{Julian McAuley}.}
  \bibinfo{year}{2020}\natexlab{}.
\newblock \showarticletitle{Attentive Sequential Models of Latent Intent for
  Next Item Recommendation}.
\newblock \bibinfo{journal}{\emph{Proceedings of The Web Conference 2020}}
  (\bibinfo{year}{2020}).
\newblock


\bibitem[\protect\citeauthoryear{van~der Maaten and Hinton}{van~der Maaten and
  Hinton}{2008}]%
        {Maaten2008VisualizingDU}
\bibfield{author}{\bibinfo{person}{Laurens van~der Maaten} {and}
  \bibinfo{person}{Geoffrey~E. Hinton}.} \bibinfo{year}{2008}\natexlab{}.
\newblock \showarticletitle{Visualizing Data using t-SNE}.
\newblock \bibinfo{journal}{\emph{Journal of Machine Learning Research}}
  \bibinfo{volume}{9} (\bibinfo{year}{2008}), \bibinfo{pages}{2579--2605}.
\newblock


\bibitem[\protect\citeauthoryear{Vaswani, Shazeer, Parmar, Uszkoreit, Jones,
  Gomez, Kaiser, and Polosukhin}{Vaswani et~al\mbox{.}}{2017}]%
        {Vaswani2017AttentionIA}
\bibfield{author}{\bibinfo{person}{Ashish Vaswani}, \bibinfo{person}{Noam~M.
  Shazeer}, \bibinfo{person}{Niki Parmar}, \bibinfo{person}{Jakob Uszkoreit},
  \bibinfo{person}{Llion Jones}, \bibinfo{person}{Aidan~N. Gomez},
  \bibinfo{person}{Lukasz Kaiser}, {and} \bibinfo{person}{Illia Polosukhin}.}
  \bibinfo{year}{2017}\natexlab{}.
\newblock \showarticletitle{Attention is All you Need}.
\newblock \bibinfo{journal}{\emph{ArXiv}}  \bibinfo{volume}{abs/1706.03762}
  (\bibinfo{year}{2017}).
\newblock


\bibitem[\protect\citeauthoryear{Wu, Li, Hsieh, and Sharpnack}{Wu
  et~al\mbox{.}}{2020}]%
        {Wu2020SSEPTSR}
\bibfield{author}{\bibinfo{person}{Liwei Wu}, \bibinfo{person}{Shuqing Li},
  \bibinfo{person}{Cho-Jui Hsieh}, {and} \bibinfo{person}{James Sharpnack}.}
  \bibinfo{year}{2020}\natexlab{}.
\newblock \showarticletitle{SSE-PT: Sequential Recommendation Via Personalized
  Transformer}.
\newblock \bibinfo{journal}{\emph{Fourteenth ACM Conference on Recommender
  Systems}} (\bibinfo{year}{2020}).
\newblock


\bibitem[\protect\citeauthoryear{Yan, Dong, Gao, Fu, Zhao, Sun, and
  McAuley}{Yan et~al\mbox{.}}{2022}]%
        {yan22personalized}
\bibfield{author}{\bibinfo{person}{An Yan}, \bibinfo{person}{Chaosheng Dong},
  \bibinfo{person}{Yan Gao}, \bibinfo{person}{Jinmiao Fu},
  \bibinfo{person}{Tong Zhao}, \bibinfo{person}{Yi Sun}, {and}
  \bibinfo{person}{Julian McAuley}.} \bibinfo{year}{2022}\natexlab{}.
\newblock \showarticletitle{Personalized complementary product recommendation}.
  In \bibinfo{booktitle}{\emph{WWW}}.
\newblock


\bibitem[\protect\citeauthoryear{Yuan, Karatzoglou, Arapakis, Jose, and
  He}{Yuan et~al\mbox{.}}{2019}]%
        {Yuan2019ASC}
\bibfield{author}{\bibinfo{person}{Fajie Yuan}, \bibinfo{person}{Alexandros
  Karatzoglou}, \bibinfo{person}{Ioannis Arapakis}, \bibinfo{person}{Joemon~M.
  Jose}, {and} \bibinfo{person}{Xiangnan He}.} \bibinfo{year}{2019}\natexlab{}.
\newblock \showarticletitle{A Simple Convolutional Generative Network for Next
  Item Recommendation}.
\newblock \bibinfo{journal}{\emph{Proceedings of the Twelfth ACM International
  Conference on Web Search and Data Mining}} (\bibinfo{year}{2019}).
\newblock


\bibitem[\protect\citeauthoryear{Zhang, Zhao, Liu, Sheng, Xu, Wang, Liu, and
  Zhou}{Zhang et~al\mbox{.}}{2019}]%
        {Zhang2019FeaturelevelDS}
\bibfield{author}{\bibinfo{person}{Tingting Zhang}, \bibinfo{person}{Pengpeng
  Zhao}, \bibinfo{person}{Yanchi Liu}, \bibinfo{person}{Victor~S. Sheng},
  \bibinfo{person}{Jiajie Xu}, \bibinfo{person}{Deqing Wang},
  \bibinfo{person}{Guanfeng Liu}, {and} \bibinfo{person}{Xiaofang Zhou}.}
  \bibinfo{year}{2019}\natexlab{}.
\newblock \showarticletitle{Feature-level Deeper Self-Attention Network for
  Sequential Recommendation}. In \bibinfo{booktitle}{\emph{IJCAI}}.
\newblock


\end{thebibliography}


\end{document}